\documentclass[aps,pra,twocolumn,superscriptaddress,showpacs,floatfix]{revtex4-1}

\usepackage{hyperref}
\usepackage{cleveref}	
\usepackage{amsmath}
\usepackage{txfonts}
\usepackage{microtype}
\usepackage{graphicx}
\usepackage{color}
\usepackage{ulem}

\begin{document}

\author{Pei-Lun He}\thanks{\mbox{P.-L.H. and M.K. contributed equally to this work.}}
\affiliation{Max-Planck-Institut f\"ur Kernphysik, Saupfercheckweg 1, 69117 Heidelberg, Germany}
\affiliation{Key Laboratory for Laser Plasmas of Ministry of Education and Department of
 Physics and Astronomy, Collaborative Innovation Center of IFSA, Shanghai Jiao Tong University, Shanghai 200240, China}
\author{Michael Klaiber}\thanks{\mbox{P.-L.H. and M.K. contributed equally to this work.}}
\affiliation{Max-Planck-Institut f\"ur Kernphysik, Saupfercheckweg 1, 69117 Heidelberg, Germany}
\author{Karen Z. Hatsagortsyan}\thanks{\mbox{Corresponding author: k.hatsagortsyan@mpi-k.de }}
\affiliation{Max-Planck-Institut f\"ur Kernphysik, Saupfercheckweg 1, 69117 Heidelberg, Germany}
\author{Christoph H. Keitel}
\affiliation{Max-Planck-Institut f\"ur Kernphysik, Saupfercheckweg 1, 69117 Heidelberg, Germany}

\bibliographystyle{apsrev4-1}

\title{Origin of high energy enhancement of photoelectron spectra in  tunneling ionization }

\date{\today}

\begin{abstract}

Recently, in a strong Coulomb field regime of tunneling ionization an unexpected large enhancement of photoelectron spectra  due to the Coulomb field of the atomic core
has been identified by numerical solution of  time-dependent Schr\"odinger equation [Phys. Rev. Lett. \textbf{117}, 243003 (2016)] in the upper energy range of the tunnel-ionized direct electrons.  We investigate the origin of the enhancement employing a classical theory with  Monte Carlo simulations of trajectories, and a quantum theory of Coulomb-corrected strong field approximation  based on the generalized eikonal approximation for the continuum electron. Although the quantum effects at recollisions with a small impact parameter yield an overall enhancement of the spectrum relative to the classical prediction, the high energy enhancement itself is shown to have a classical nature and is due to momentum space bunching  of photoelectrons released not far from the peak of the laser field. The bunching is  caused by a large and nonuniform, with respect to the ionization time, Coulomb momentum transfer at the ionization tunnel exit.

\end{abstract}

\maketitle

\textit{Introduction.} The Coulomb field of the atomic core plays a significant role for strong field ionization. Since long time it has been known that it lowers the tunneling barrier and increases the tunneling probability \cite{Perelomov_1966b,Perelomov_1967a,Popov_1967,Popov_2004u}. Significant Coulomb effects arise at recollisions \cite{Corkum_1993}. While hard recollisions induce well-known processes of above-threshold ionization \cite{Becker_2002}, high-order harmonic generation \cite{Agostini_2004}, and  nonsequential double ionization \cite{Becker_2012}, the soft recollisions bring about  Coulomb focusing \cite{Brabec_1996,Yudin_2001a,Comtois_2005} and defocusing \cite{Kelvich_2016} effects. The Coulomb focusing is responsible for the, so-called, low-energy structures (LES) \cite{Blaga_2009,Catoire_2009,Quan_2009,Wu_2012b,Wolter_2014,Dura_2013,
Pullen_2014,Xia_2015,Wolter_2015x,Zhang_2016,Diesen_2016,Williams_2017,
Liu_2010,Yan_2010,Kastner_2012,Lemell_2012,Becker_2015}, especially conspicuous and lately discovered in mid-IR laser fields.

Recently, another surprising Coulomb field effect has been identified by ab initio numerical solution of  time-dependent Schr\"odinger equation (TDSE) \cite{Keil_2016}. When comparing the numerical solution  for the photoelectron spectra with calculations of the  first-order Coulomb-free
strong field approximation (SFA) \cite{Keldysh_1965,Faisal_1973,Reiss_1980}, several orders enhancement of photoelectron spectra at $2U_p$, i.e., twice of the electron ponderomotive energy, has been observed. In \cite{Keil_2016} the effect was analyzed invoking the Coulomb-corrected action along  quantum orbits in the complex-time plane. Due to the Coulomb field, the quantum orbit maintains a large imaginary part up to the recollision, which hinted a  conclusion that the enhancement  is a specific quantum effect, and that separation into sub-barrier motion up to the  tunnel exit  and subsequent classical motion is an invalid concept. Although, the high energy enhancement in \cite{Keil_2016} is traced back to the Coulomb field effect,  an intuitive understanding  remained missing.

The aim of this Letter is to clarify the origin of the high energy Coulomb enhancement (HECE) in the photoelectron spectrum. We carry out a classical as well as a quantum mechanical analysis. The classical analysis employs the classical trajectory Monte Carlo (CTMC) simulations with nonadiabatic initial conditions for the electrons. For the quantum mechanical analysis we put forward a new version of Coulomb-corrected strong field approximation (CCSFA). In the existing theories of CCSFA, such as the Trajectory-based CCSFA \cite{Popruzhenko_2008a,Popruzhenko_2008b}, or Analytical R-matrix theory \cite{Torlina_2012,Torlina_2012b,Kaushal_2013}, the Coulomb field of the atomic core is accounted for using the eikonal wave function for the continuum electron. In the latter Wentzel-Kramers-Brillouin (WKB) approximation is applied, with a perturbative treatment of the Coulomb potential in the phase of the wave function. Unfortunately, the eikonal CCSFA has a singularity for the forward re-scattering amplitude, cf. \cite{Keil_2016}, which renders the HECE treatment ambiguious. We go beyond the WKB description of the continuum electron,  incorporating into the SFA formalism the electron wave function in the, so-called, generalized eikonal approximation (GEA) \cite{Kaminski_1984,Avetissian_1997,Avetissian_1999,Velez_2015}. In GEA the second order derivatives of the Schr\"odinger equation are not neglected, in contrast to WKB approximation. The latter allows to take into account quantum recoil effects at recollisions with a small impact parameter and to remove the Coulomb singularity of the eikonal CCSFA at recollisions.  The accuracy of our analytical results are examined in comparison with numerical solutions of TDSE.

\begin{figure}
   \begin{center}
    \includegraphics[width=0.25\textwidth]{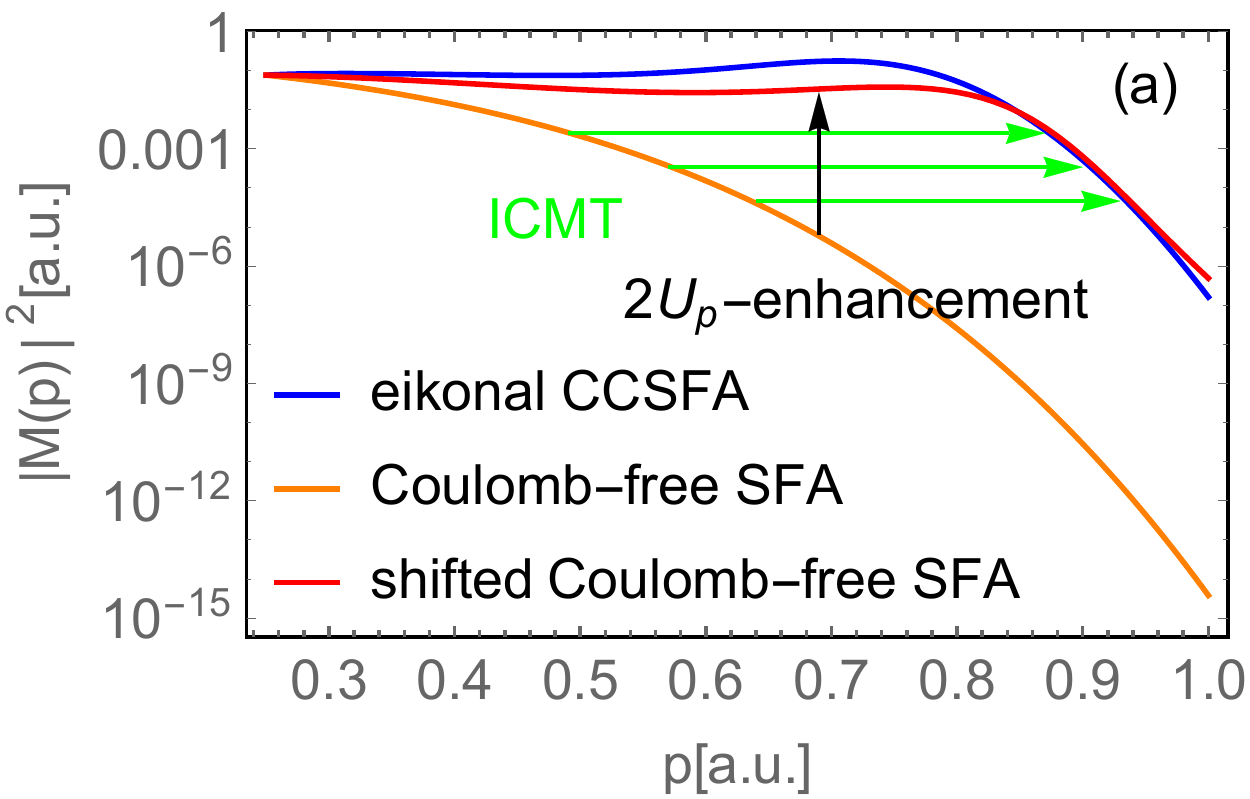}
\includegraphics[width=0.22\textwidth]{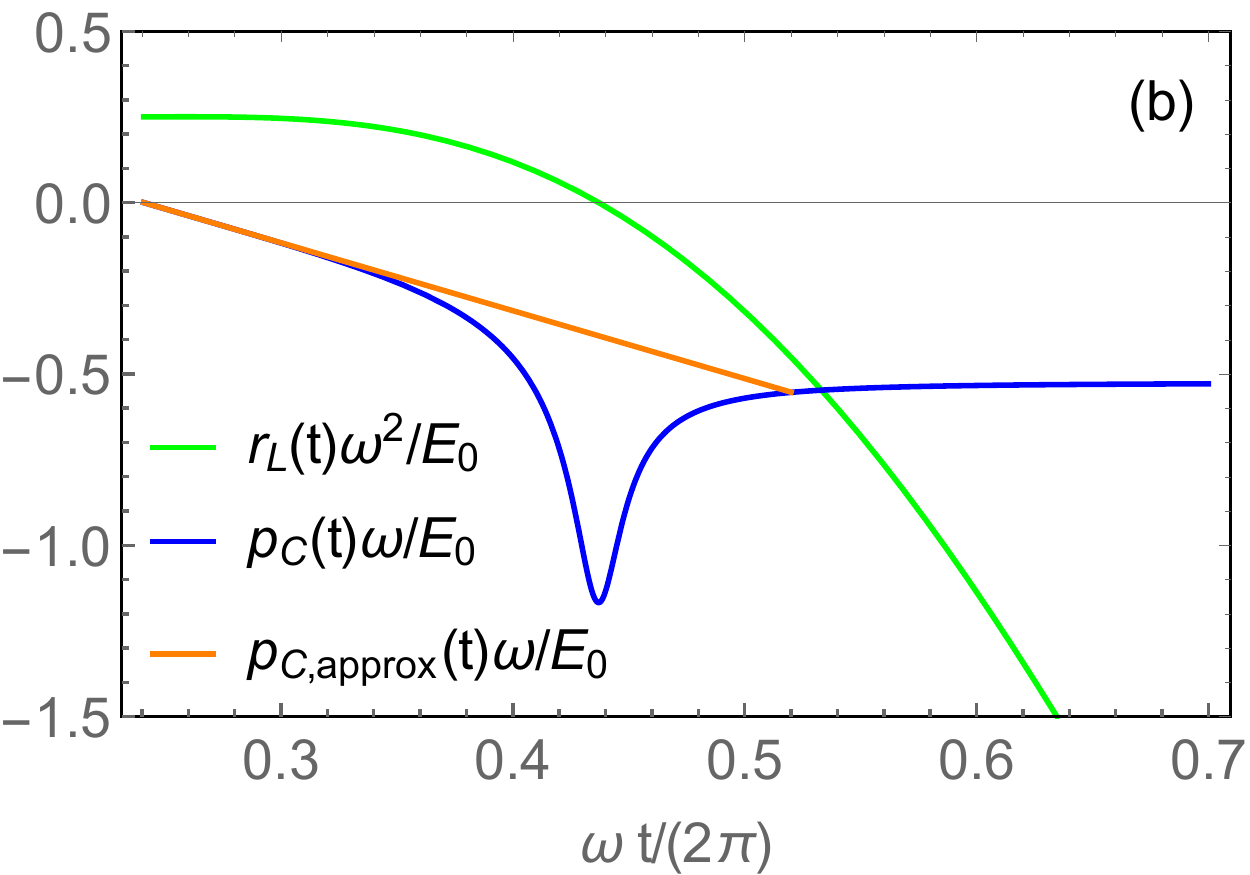}
           \caption{(a) Photoelectron momentum distribution at a fixed emission angle $\theta=\pi/100$: (red) via $|M^C_{\textbf{p}}|^2$ corresponding to the long trajectory; (blue) via eikonal CCSFA as in \cite{Keil_2016}; (orange) the Coulomb free 1st-order SFA. The shifts of the momentum distribution due to the initial Coulomb momentum transfer (ICMT) are shown by green arrows, and HECE is indicated by a black arrow. (b)   Coulomb momentum transfer along the laser field $p_{Cx}(t)/(E_0/\omega)$ vs the interaction time, at $p=0.69$ and $\theta=\pi/100$: (blue) numerical evaluation, (orange) estimation, see the text; (green) the electron trajectory $x(t)/(E_0/\omega^2)$. The parameters are $E_0=0.0045$ a.u., $\omega=0.0065$ a.u., $I_p=0.14$ a.u.,  and $Z=1$. }
       \label{MCp}
    \end{center}
  \end{figure}

Firstly, we show that already CTMC simulations with nonadiabatic initial conditions reproduce qualitatively  the HECE. We analyze the trajectories yielding high energies and trace the origin of the enhancement. It is due to electrons released not far from the peak of the laser field, though with bunching at high energies because of a large, nonuniform Coulomb momentum transfer,
which depends on the ionization phase, i.e., the laser phase at the ionization. In contrast to LES, here we deal mostly with Coulomb defocusing, and the enhancement is not due to Coulomb focusing.
Although HECE is mainly determined by the parameter $\zeta\equiv Z\omega/E_0 > 1$, pointed out already  in \cite{Keil_2016}, with the frequency $\omega$ and  the amplitude $E_0$ of the laser field, we find an additional dependence on the ionization potential $I_p$. The enhancement is larger with smaller $I_p$, which however is reverted at very small $I_p$. The latter is found to be related to restructuring of the topological structure of the initial phase space of the ionized electrons when approaching  the regime of  over-the-barrier ionization (OTBI). Secondly, the applied quantum approach with GEA allows us to remove the Coulomb singularity of the eikonal CCSFA at recollisions, and to obtain a reliable quantum description for the photoelectron spectra near the upper energy limit of the direct electrons and, consequently, for HECE. The quantum description induces merely a uniform enhancement of the photoelectron spectra compared to the classical result, i.e., the nature  of HECE is classical.

\textit{Qualitative discussion.} Before going to rigorous calculations, we  illustrate the origin of HECE  with the following qualitative discussion. We begin with the 1st-order  SFA amplitude describing the tunnel ionized direct electrons, neglecting the  Coulomb effect of the atomic core: $M_\mathbf{p}\sim \sum_s \exp\left(-i\int_{t_s}dt(\mathbf{p}+\mathbf{A}(t))^2/2+iI_pt_s\right)$, where $\mathbf{p}$ is the final momentum, $\mathbf{A}(t)=\mathbf{e}_x(E_0/\omega)\sin(\omega t)$ is the  vector potential of the linearly polarized laser field, and $t_s$ is the time-saddle point of the relevant trajectory. Then, we derive the Coulomb corrected photoelectron momentum distribution by means of nonuniform (depending on the ionization phase) momentum shifting of the 1st-order momentum distribution: $M^C_\mathbf{p} = M_{\mathbf{p}- \mathbf{p}_C}$, where $\mathbf{p}_C=-Z\int^\infty_{t_e}dt\,\mathbf{r}_L(t)/r_L^3(t)$ is the Coulomb momentum transfer to the electron along the laser driven trajectory $\mathbf{r}^L(t)$, $r_L(t)=|\mathbf{r}_L(t)|$, $t_e={\rm Re}\{t_s\}$ is the ionization time and $x_e={\rm Re}\left\{{\int^{t_e}_{t_s}dt(p_x+A(t))}\right\}$ is the tunnel exit coordinate. For the short trajectory $p_{xC}$ is opposite to the  the final longitudinal momentum $p_x$, while for the long trajectory they have the same sign. Accordingly, the Coulomb momentum shift increases the probability $|M^C_\textbf{p}|^2$ for the long trajectory, because the electron with a certain final momentum is ionized closer to the laser peak than in the Coulomb free case, and vice verse for the short trajectories. The photoelectron energy distribution via $|M^C_\textbf{p}|^2$ for recolliding long trajectories is shown in Fig.~\ref{MCp}(a).
The photoelectron spectrum  demonstrates a plateau-like behavior up to $2U_p$-energy, similar to the result of  \cite{Keil_2016}, and indicates that HECE arises due to the nonuniform Coulomb momentum transfer to the continuum electron along real trajectories. In this case the electrons with final energy around $2U_p$  are tunneled from the atom not near the zero crossing of the laser field, as in the Coulomb free case, but at the laser phases close to the peak of the field with  enhanced probabilities.

\begin{figure}
\centering
\includegraphics[width=0.5\textwidth]{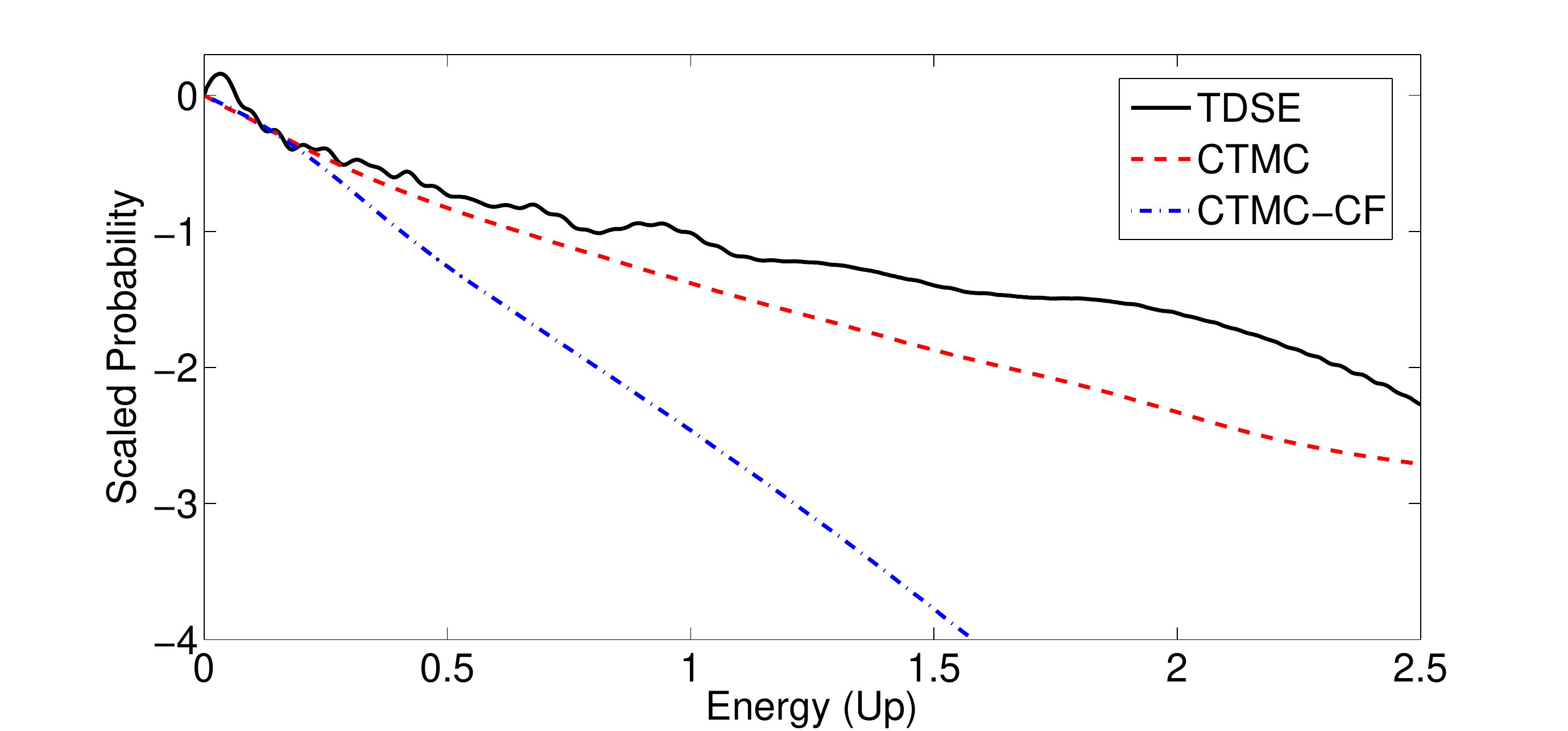}
\caption{\label{CTMC} Photoelectron spectra: (black-solid) numerical solution of TDSE,  (red-dashed) CTMC, (blue-dot-dashed) Coulomb free CTMC. The laser and atom parameters are as in Fig.~\ref{MCp}.  }
\end{figure}
\begin{figure}[b]
\centering
\includegraphics[width=0.5\textwidth]{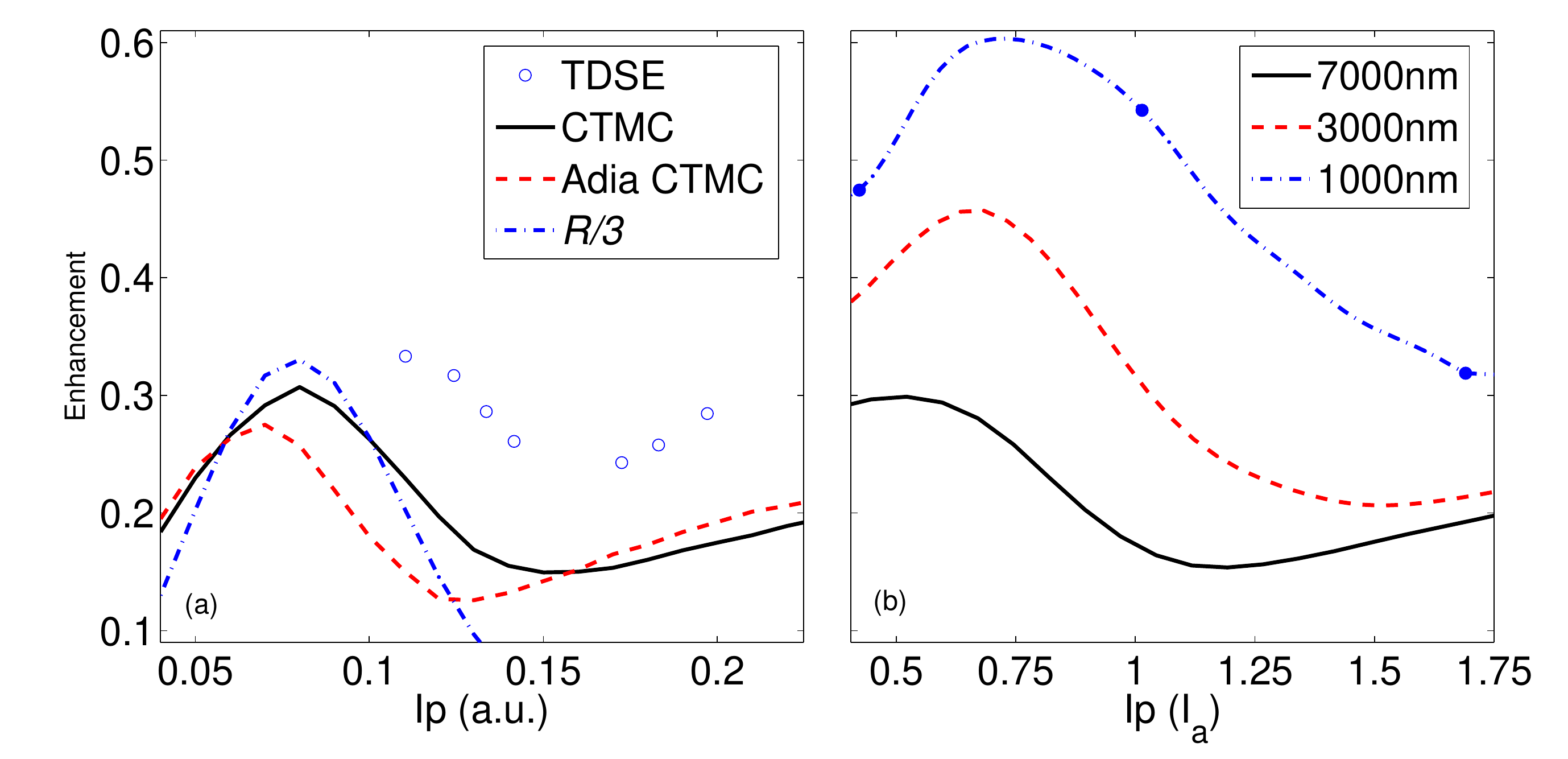}
\caption{The ratio ${\cal R}$ of ionization probabilities at $2U_p$ to $U_p$: (a) vs $I_p$ at $E_0=0.0045$, $\omega=0.0065$ a.u.  and $Z=1$, (black-solid) CTMC nonadiabatic, (red-dashed) CTMC adiabatic, (cycles) TDSE, (dot-dashed) scaled estimation for ${\cal R}$, see the text; (b) vs $I_p/I_a$ for different wavelengths at $Z\omega/E_0=1.44$, $I_a=\sqrt{4ZE_0}$. The dots on panel (b) correspond to three cases with different $I_p$s applied in Fig.~\ref{inphase}.}
\label{RRRR}
\end{figure}
 \begin{figure*}
\centering
\includegraphics[width=0.9\textwidth]{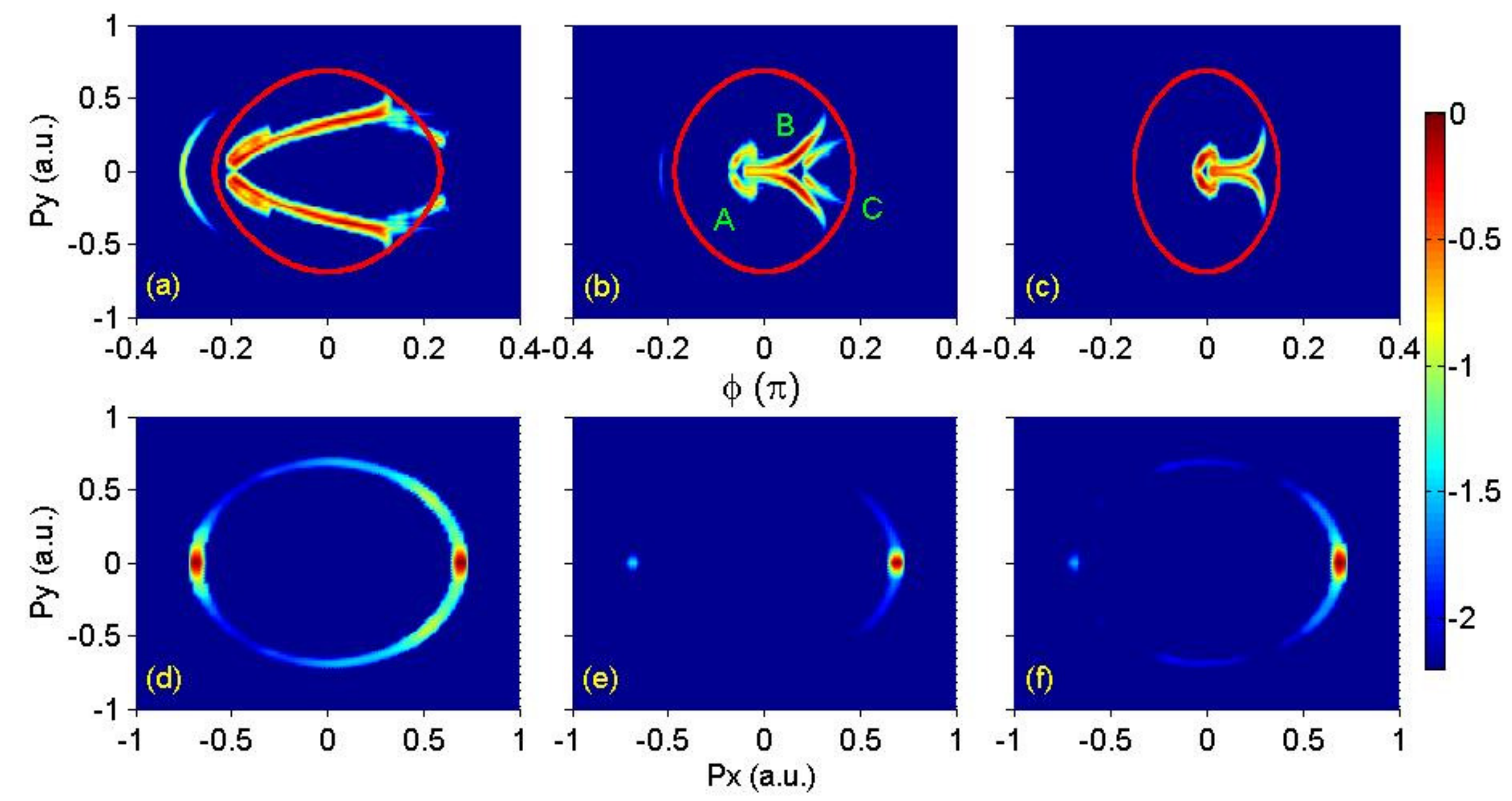}
\caption{ The electron initial phase space ($p_{\bot i},\phi_i$) with color coded probabilities, which finally contributes to the photoelectron energy interval ($1.9 U_p,2.1 U_p$) (first row). The red elliptic regions correspond to the Coulomb free case. Asymptotic momentum distribution (second row). The phase space of the trajectories A, B, and C type  are indicated in the panel (b).  (a), (d) $I_p=0.42 I_a$ (before the HECE peak);
(b), (e) $I_p=1.01 I_a$ (near the HECE peak);
(c), (f) $I_p=1.69 I_a$  (after the HECE peak); these cases are indicated by dots in Fig.~\ref{RRRR}(b).
 The parameters are $E_0=0.0315$ a.u., $\omega=0.0456$ a.u. and $Z=1$. }
\label{inphase}
\end{figure*}

We can estimate the scaling for HECE analyzing the relevant trajectories. The long  laser driven trajectory at the $2U_p$-cutoff is launched near zero crossing of the field and the electron is initially   almost standing still, further moving along the approximate trajectory $x(t)\approx x_e-E_0 \omega (t-t_e)^3/6$ \cite{Suppl_material} [see the trajectory (green line) in Fig.~\ref{MCp}(b)], which admits a simple estimate for the Coulomb momentum transfer:
$p_{C} \sim Z\delta t/x_e^2\sim 4Z (E_0/E_a) (3/\gamma)^{1/3}\propto I_p^{-5/3}$ \cite{Suppl_material}, see Fig.~\ref{MCp}(b). In the latter, the effective time interval $\delta t$ is estimated by the time the electron covers the distance $|x_e|\sim I_p/E_0$, and $E_a\equiv (2I_p)^{3/2}$ is the atomic field.
HECE  can be estimated as ${\cal E}\equiv \left.|M^C_{p}|^2/|M_{p}|^2\right|_{p=p_0}$, $p_0=E_0/\omega$, using the 1st-order SFA longitudinal momentum distribution $|M_p|^2\sim \exp(-p^2_\Vert/\Delta_{\Vert}^2)$, with $\Delta_\Vert=\sqrt{3E_0/E_a}E_0/\omega$ \cite{Popov_2004u}. Thus, we find ${\cal E}\sim \exp (4\overline{\zeta })$ \cite{Suppl_material}, with the enhancement factor
\begin{eqnarray}
\overline{\zeta }\approx
\frac{Z\omega}{E_0}\frac{1}{\gamma^{1/3}},
\label{zeta}
\end{eqnarray}
where $\gamma=\omega\sqrt{2I_p}/E_0$ is the Keldysh parameter  \cite{Keldysh_1965}. The ratio ${\cal R}$ of ionization probabilities at $2U_p$ to $U_p$ can be regarded as a signature of HECE to be proved in an experiment ($U_p$ is used as a reference point to avoid the spikes in the spectrum due to LES).
It can be estimated as ${\cal R}= |M^C_{p_0}|^2/|M_{p_1}|^2 \approx e^{\left[(p_1-p_{C1})^2-(p_0-p_C)^2\right]/\Delta_\Vert^2}$,
with  $p_1=E_0/(\sqrt{2}\omega)$, and $p_{C1}=\pi ZE_0/E_a$ \cite{Shvetsov-Shilovski_2009}, which is illustrated in Fig.~\ref{RRRR}(a).

\textit{Classical description.} To corroborate the classical nature of HECE, we have carried out CTMC simulations with nonadiabatic initial conditions \cite{Li_2016}. One can deduce from Fig.~\ref{CTMC}  that the classical simulation
shows already the enhancement stemming from the Coulomb field effect and fits qualitatively to the numerical solution of TDSE. Nonadiabatic initial conditions are favorable for the enhanced Coulomb effect, because the  tunnel exit in this case is closer to the core and an initial longitudinal momentum appears which increases the time the electron spent near the tunnel exit.

Further, we analyze the dependence of the enhancement on the laser and atom parameters,
see Fig.~\ref{RRRR}. As a characteristic of the enhancement we use the ratio ${\cal R}$ of the probability at energy $2U_p$ to that at the energy $U_p$. While the parameter $\overline{\zeta}$ qualitatively describes the decreasing behavior of HECE with moving $I_p$ far from the peak. However, we find that the enhancement additionally depends  on the laser wavelength, and crucially on the ionization potential. The HECE is peaked  at around $I_p\approx I_a=\sqrt{4ZE_0}$ a.u., when the transition to OTBI takes place \cite{Augst_1989} and the  enhancement character  qualitatively changes.

For intuitive understanding of the enhancement mechanism we investigate the initial momentum space ($p_{\bot i},\phi_i$) of the trajectories that contribute to HECE within the final energy interval of $(1.9U_p, 2.1U_p)$, see Fig.~\ref{inphase}. In the Coulomb free case  the contribution to the $2U_p$ energy is not large because either the initial transverse momentum $p_{\bot i}$ is large, or the ionization phase is far from the peak value $\phi_i=0$, see the red ellipse in the first row of Fig.~\ref{inphase}.
In contrast to that, when the Coulomb field is accounted for, the electrons contributing to the  final $2U_p$ energy range are ionized with smaller $p_{\bot i}$ and $\phi_i$ (i.e. closer to the peak of the field) with enhanced ionization probabilities.

The typical parameter regime of HECE corresponds to Fig.~\ref{inphase}(b),(e).  The most of HECE contribution comes from trajectories B (a typical trajectory is shown in Fig.~\ref{MCp}(b)). Moving along the initial phase structure of B from small values of $\phi_i$ and $p_{\bot i}$ to the larger ones,  corresponds to transition from the wings of the final $2U_p$ energy ring to the central spot at $p_{\bot }=0$ in Fig.~\ref{inphase}(e) \cite{Suppl_material}. For the former, the final large energy is achieved due to a large transverse Coulomb momentum transfer at a recollision, while for the latter due to an initial Coulomb momentum transfer at the tunnel exit. In this parameter regime the densities of the initial phase space for the trajectories of the type A and C are small.

When increasing $I_p$  the Coulomb momentum transfer diminishes, $p_C\propto I_p^{-5/3}$, and the initial phase-space of trajectories B moves far from $\phi_i=0$, cf.  Fig.~\ref{inphase} (b) and (c). In this case the contribution of trajectories A  is increased with respect to B. The contribution of trajectories C becomes negligible. The trajectories A are chaotic and the total initial phase space of contribution electrons is decreased, with a result of decreasing HECE, see Fig.~\ref{RRRR}.

The maximum of the enhancement is achieved when the phase-space of trajectories B and C merge at decreasing $I_p$ \cite{Suppl_material}.
At the further decrease of $I_p$, see Fig.~\ref{inphase}(a), the phase-space of all trajectories A, B, and C are merged and  the topological structure of the initial phase-space is altered \cite{Suppl_material}. This results in the increase of $p_{\bot i}$, which again suppresses HECE. The alteration of the structure of the initial phase-space at decreasing the ionization potential is related to the transition of ionization from the tunneling to the over-the-barrier ionization.

While in LES the enhancement is due to Coulomb focusing, in HECE this plays minor role. We classify the trajectories as Coulomb focused if $|p_{\bot i}|>|p_{\bot f}|$, otherwise  Coulomb defocused. In the first case the asymptotic transverse momentum space of the ionized electron is shrunk  with respect to that at the tunnel exit, which leads to an additional enhancement \cite{Brabec_1996}.
The weight of defocused trajectories contributing to HECE is larger than 70\% \cite{Suppl_material}.

\begin{figure}
   \begin{center}
 \includegraphics[width=0.5\textwidth]{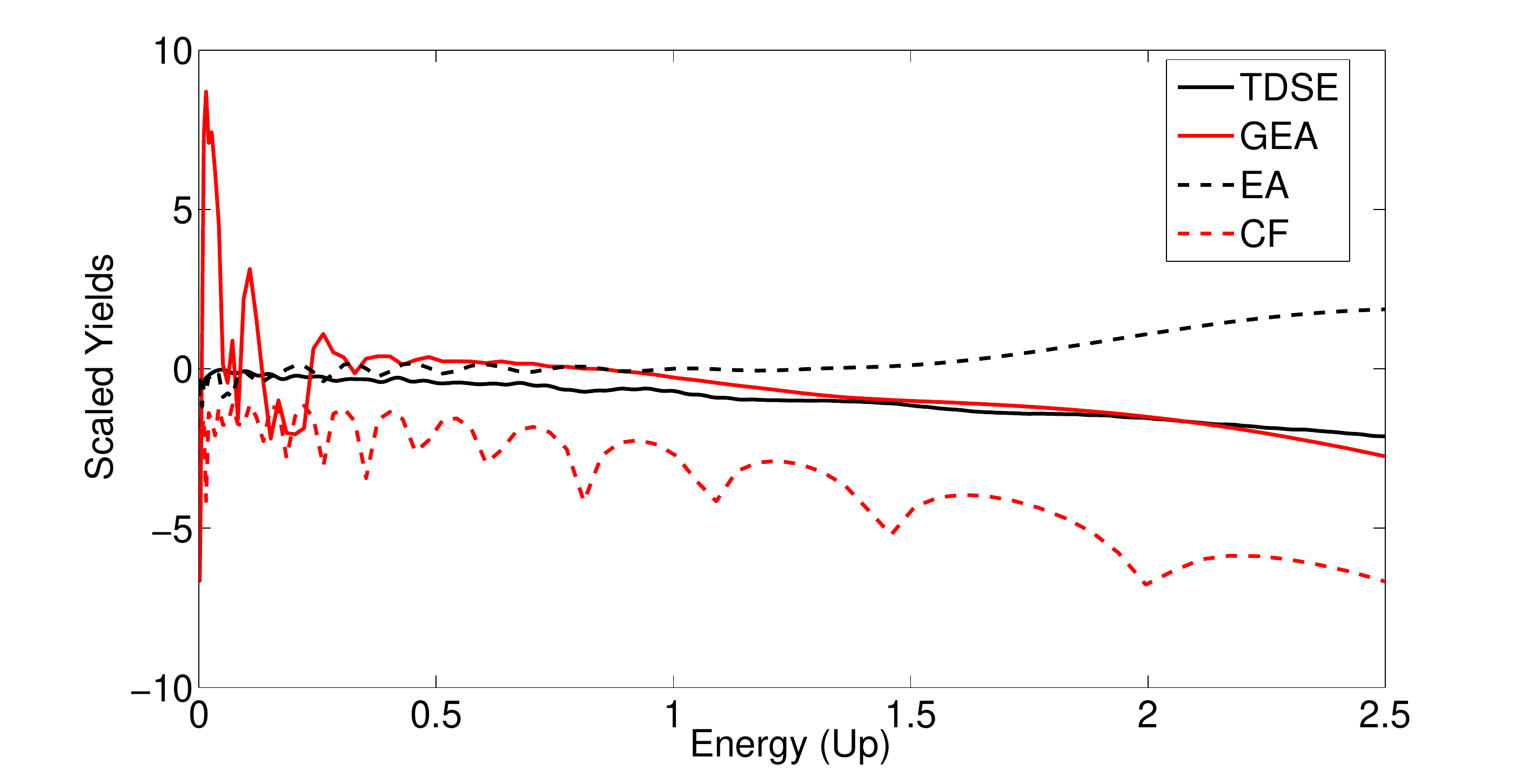}
  \caption{Photoelectron  spectra angle-integrated within $\pm 6^0$ along the laser polarization direction: (black-solid) via numerical TDSE; (red-solid) via GEA; (blue-thick-dashed) via eikonal CCSFA as in \cite{Keil_2016}; (black-thin-dashed) via Coulomb free SFA. The laser and atom parameters are the same as in Fig.~\ref{CTMC}.}
       \label{GEAspectrum}
    \end{center}
  \end{figure}

\textit{Quantum description. Generalized eikonal approach.}
For the description of HECE a nonperturbative treatment of the Coulomb effect is necessary because the perturbative second order SFA yields uniformly enhanced photoelectron spectra, while at HECE the enhancement is large at high energies around $2U_p$ \cite{Suppl_material}. In \cite{Keil_2016} CCSFA is applied which employs eikonal approximation for the electron continuum wave function. The deficiency of this approach is that the ionization amplitude diverges at photoelectron rescattering to small angles, which induces an artificial large contribution to the photoelectron spectra enhancement at high energies , see Fig.~\ref{GEAspectrum}. We remedy the divergence problem at recollisions using the generalized eikonal wave function in the CCSFA approach, which includes quantum corrections.

The photoelectron momentum distribution in CCSFA is calculated via the following matrix element:
\begin{eqnarray}
M_\mathbf{p}=-i\int dt d^3\mathbf{r}\, \psi^{GEA\,*}(\mathbf{r},t)\,\mathbf{r}\cdot\mathbf{E}(t)\,\phi(\mathbf{r},t),\label{Mp}
\end{eqnarray}
where the electron wave function in the continuum $\psi^{GEA}(\mathbf{r},t)$ accounts for the effect of the laser and Coulomb field of the atomic core in the generalized eikonal approximation \cite{Velez_2015}:
\begin{eqnarray}
\psi^{GEA}(\mathbf{r},t)=\frac{1}{\sqrt{\det \partial \mathbf{p}_f/\partial \mathbf{p}_i}}
\exp \left[i S_0(\mathbf{r},t)+i S^{GEA}(\mathbf{r},t)\right],\label{wf_GEA}
\end{eqnarray}
with the Volkov action $S_0=\int_t dt' (\textbf{p}+\textbf{A}(t'))^2/2+[\textbf{p}+\textbf{A}(t)]\cdot \textbf{r}$, and the generalized eikonal $S^{GEA}(\mathbf{r},t)\approx S^{GEA}_1(\mathbf{r},t)+S^{GEA}_2(\mathbf{r},t)$. Here the prefactor originates from the wave function normalization and describes the Jacobian of the momentum space transformation. We calculate the generalized eikonal up to the second order in the scattering potential $V(\textbf{r})$ \cite{Suppl_material}:
\begin{eqnarray}
S^{GEA}_1(\mathbf{r},t)=\int_t ds V(\mathbf{r}_L(s))\rm{erf}\left[\sqrt{\frac{i\mathbf{r}_L(s)^2}{2(s-t)}}\right].
\end{eqnarray}
and
\begin{eqnarray}
S^{GEA}_2(\mathbf{r},t)=\frac{1}{2}\int_t ds \left(\int_s ds' \boldsymbol{\nabla}V(\mathbf{r}_L(s'))\rm{erf}\left[\sqrt{\frac{i\mathbf{r}_L(s')^2}{2(s'-t)}}\right]\right)^2,
\end{eqnarray}
with the laser driven trajectory $\mathbf{r}_L(s)$. We solve the integrals in the amplitude of Eq.~(\ref{Mp})  with the saddle point method and expand the saddle points  up to first order in the atomic potential. The photoelectron spectrum along the laser polarization in GEA is presented in Fig.~\ref{GEAspectrum}. It shows enhancement with respect to the Coulomb-free case, and the same slope for the spectrum asfor the TDSE result. The  Coulomb-focusing accounted for via the prefactor determinant in Eq.~(\ref{wf_GEA}) has  little contribution in this enhancement \cite{Suppl_material}. The difference between the quasiclassical eikonal approximation and the GEA occurs at 2$U_p$ energies where the perturbative quasiclassical approximation is not valid. The GEA overestimates the LES, see qualitative estimation in  \cite{Suppl_material}.

\textit{Conclusion.} we have demonstrated that the enhancement of the tunnel ionized photoelectron spectra at the upper energy limit of the direct electrons in the strong Coulomb field regime is of a classical origin. We found that the nonuniform Coulomb momentum transfer with respect to the ionization phase allows for the electrons tunneled not far from the peak of the laser field to accumulate at high energies.
The enhancement not only depends on the main parameter of the strong Coulomb field regime $Z\omega/E_0$, but also crucially on the ionization potential. We locate a peak of the enhancement  with respect to the ionization potential and relate this to the structure of the initial phase-space of the contributing electrons. In contrast to LES, the Coulomb focusing plays no role for HECE. For the accurate quantum description of the Coulomb effects at fast recollisions, we put forward a new Coulomb corrected version of SFA based on the generalized eikonal approximation which is free from  Coulomb divergence at recollisions.

P.-L.H. acknowledges the support from the China Scholarship Council (CSC).

\bibliography{strong_fields_bibliography}

\end{document}